\documentclass[structabstract,printer]{aa} 
\usepackage{graphicx}
\usepackage{natbib}
\usepackage{amsmath}
\usepackage{txfonts}
\usepackage{color}
\usepackage{enumerate}

\newcommand{\be}{\begin{eqnarray}}
\newcommand{\ee}{\end{eqnarray}}

\newcommand {\nbodypp}{\textsc{\mbox{nbody6\raise.4ex\hbox{\tiny++}}}}
\newcommand {\Msun} {\mbox{M$_{\odot}$}}

\begin{document}

\title{Which young clusters/associations are  we missing today?}
\author{Susanne Pfalzner, Kirsten Vincke, and Mai Xiang}
\institute{
\inst{1}Max-Planck-Institut f\"ur Radioastronomie, Auf dem H\"ugel 69, 53121 Bonn, Germany\\
\email{spfalzner@mpifr.de}}
\date{ }

\titlerunning{Observable clusters in the 5-30 Myr range}
\authorrunning{Susanne Pfalzner}

\abstract{Currently clusters/associations of stars are mainly detected as surface density enhancements relative to the background field. While clusters form, their surface density increases. It likely decreases again at the end of the star formation process when the system expands as a consequence of gas expulsion. Therefore the surface density of a single cluster can change considerably in young clusters/associations during the first 20 Myr of their development.}
{We investigate the effect of the gas expulsion on the detectability of clusters/associations typical for the solar neighbourhood, where the star formation efficiency is $<$35\%. The main focus will be laid on the dependence on the initial cluster mass.}
{ Nbody methods are used to determine the cluster/association dynamics after gas expulsion.}
{We find that, even for low background densities, only clusters/associations  with initial central surface densities exceeding a few 5000 \Msun/pc$^2$ will be detected as clusters at ages $\gtrapprox$ 5 Myr.  Even the Orion Nebula cluster, one of the most massive nearby clusters, would  only be categorized as a small co-moving group  with current methods after 5 Myr of development. This means that cluster expansion leads to a selection effect -  at ages of $<$1-2 Myr the full range of clusters/associations is observed whereas at ages $>$ 4 Myr  only the most massive clusters are identified, while systems with initially $M_c <$ 3 000 \Msun\ are missing.}
{The temporal development of stellar properties is usually determined by observing clusters of different ages.  The potentially strong inhomogeneity of the cluster sample makes this methods highly questionable. However, GAIA could provide the means to rectify this situation as it will be able to detect lower mass clusters. }

\maketitle
\section{Introduction}

Many young clusters/associations\footnote{In the following we abbreviate clusters/associations as clusters avoiding the semantic difficulty that young groups of physically-related stars are referrred to as clusters, whereas the same group might be referred to as association after gas expulsion. All exposed clusters considered here would be classified as associations or extended clusters according to the definition given in Gieles \& Portegies-Zwart (2010) or Pfalzner \& Kaczmarek (2013b), respectively.} are known that are still embedded in their natal gas and as such still actively forming stars. Such embedded clusters have typically ages of $\leq$3 Myr. For this group plenty of information concerning star and planet formation is available.
For clusters that have completed the star formation process the situation is different: just a few clusters in the age range 3-10 Myr
are known in the Milky Way, among them are for example,  NGC 2244, the Upper Scorpius association (Upp Sco), and Ori 1a.

Lada \& Lada (2003)   were the first to point out the rapid decline in the number of observed clusters after the embedded phase in the solar neighbourhood. In the Milky Way most stars form in clusters, but by an age of 10-100 Myr only a few per cent of stars remain part of clusters with stellar densities noticeably above that of the field. Whether this is the same in other galaxies is still an open question (e.g., Silva-Villa and Larsen 2011, Fall and Chandar 2012). 

Gravitationally bound clusters will not disperse on their own over this time-scale as two-body evaporation does not become important until an age of 100-1000 Myr. This is so even for a modest cluster of $N$ = 1000 (Krumholz et al. 2014). Thus, the question is what drives this rapid decline in cluster numbers over the first 10 Myr.

Currently there exist many different ideas how the extended young clusters in the solar neighbourhood evolve over the first 10 Myr after their formation and which are the key parameters in this evolution\footnote{Note that the development of the compact clusters in the Milky Way or nearby galaxies might differ considerably from what is described here (Silva-Villa et al. 2013, Pfalzner \& Kaczmarek 2013b. }Many models  expect gas expulsion to be the main driver of the dynamics during this time span (Lada \& Lada 2003, Krumholz et al. 2014),  others argue for gas expulsion being of minor importance. In the later scenario stars would form in gravitationally-unbound gas clouds.
An argument for this model is its similarity to the results of numerical simulations of star formation (Bate et al. 2003, Bonnell et al. 2003). However, there is a risk in comparing theoretical models with simulation rather than observations - every simulation inevitably contains simplifying assumptions which may influence the result (Krumholz 2014).  Most important in this context, these simulations invariably produce bound clusters (Krumholz  et al. 2014). As such they might be more representative for the very compact clusters, like Arches and NGC 3603, found in the spiral arms and close to the Galactic Centre with their relatively high star formation efficiencies (SFE) (Pfalzner 2009, Bastian 2012, Pfalzner \& Kaczmarek 2013b). These can probably survive at least 100 Myr and are therefore not representative for the short-lived clusters typically found in the solar neighbourhood. For additional critical arguments of this scenario see the recent review by Krumholz et al. (2014).

Although it cannot be excluded that stars to some extend form in gravitationally-unbound gas clouds, recent observations strongly indicate that gas expulsion plays an important role in the cluster development. For example, the Gaia-ESO survey of the $\sim$ 5Myr-old $\gamma$ Velorum cluster (Jeffries et al. 2014) shows that the cluster contains two groups of stars with distinctly different velocity dispersions, which is exactly what one would expect in the early phases of cluster expansion after gas expulsion. A similar bi-modality is found for the cluster NGC 346 in the SMC (Gouliermis et al. 2014). In addition, a recent study looking at the masses and radii of the most massive clusters in the solar neighbourhood found that their age-dependence corresponds to the temporal development of massive clusters after gas expulsion with an SFE of 30\% (Pfalzner \& Kaczmarek 2013). Given these indications for the potential importance of the gas expulsion process, we investigate here its consequences for the detectability of clusters after the gas expulsion phase.

Already Lada \& Lada (2003) cautioned that the rapid decline in the number of clusters in the age range 10 -100 Myr does not necessarily mean that these clusters dissolve completely. They might just fall below the detection limit due to mass loss and expansion. Here we want to quantify when the surface density of a cluster drops below the detection limit in a gas expulsion scenario typical for the solar neighbourhood. 

After the description of the numerical method in section 2, first the simplest case of instantaneous gas expulsion is investigated for clusters of different masses (section 3). Although probably a good approximation for massive clusters ($M_c > 10^4$ \Msun) the assumption of short gas expulsion times probably breaks down for clusters with masses $<$ 500 \Msun. Therefore we investigate the influence of the gas expulsion time on the detectability in section 4. In section 5 we discuss the consequences for the temporal development of stellar properties by investigating clusters of different ages. Here we point out as well how the observed properties of specific clusters are expected to change, when a clearer distinction between bound and unbound members will be possible with GAIA.

\begin{table}
\begin{center}
\begin{tabular}{l*{7}{c}}
ID & $N_{stars}$ & $r_c^i$ & SFE &  $t_{dur}$&  $t_{exp}$ & $N(M)$ &$N_{sim}$ \\[0.5ex]
\hline
A  1     & 30 000      & 1.3  &  0.1-0.3    & 0  &  1 &IMF        & 15\\
A  2     & 15 000      & 1.3  &    0.3         & 0  &  1 &IMF        & 30\\
A  3     &    5 000      & 1.3  &   0.3          & 0 &  1&IMF         & 90\\
A  4     &    2 000      & 1.3  &        0.3    & 0  &  1 &IMF        & 100\\
A  5     &       500      & 1.3  &   0.3          & 0 &  1, 3 &IMF    & 100\\
B        &       500      & 1.3  &   0.3          & 2 &  1 &IMF          & 400\\
\end{tabular}
\caption{Properties of the presented cluster models: ID stands for the identifier, the second column depicts the number of stars, the third column denotes the initial cluster half-mass radius in pc, the fourth column shows the range of investigated SFEs,  $t_{dur}$ stands for the duration of the gas expulsion phase in Myr,  $t_{exp}$ the time at which gas expulsion starts, $N(M)$ indicates the used mass representation and the last column the number of simulations performed for the given set-up. 
\label{table:sim}}
\end{center}
\end{table}

\section{Method}

There is a long history of modelling the cluster development after gas expulsion (Tutukov 1978, Hills 1980, Lada, Margulis  \& Dearborn 1984, Kroupa, Adams 2000, Geyer  \& Burkert 2001, Kroupa, Aarseth  \& Hurley 2001, Boily  \& Kroupa 2003, Bastian  \& Goodwin 2006,  Baumgardt \& Kroupa 2007, Smith 2011, Banerjee \& Kroupa 2014). In the context of the gas expulsion scenario, currently there is a debate about the initial conditions of clusters, specifically whether substructure is important (Smith et al. 2011) or a centrally condensed picture describes clusters better. In addition, the SFE could be constant throughout the clump or a function of the local gas density (Krumholz \& Tan 2007, Gutermuth et al. 2011, Parmentier \& Pfalzner 2013, Parmentier et al. 2014, Pfalzner et al. 2014). Here we investigate the case of a centrally concentrated clump and a constant SFE, but will perform a similar investigation for a density-dependent SFE in a follow-up study.

Another point of discussion is the virial state of a cluster before gas expulsion (Caputo et al. 2014). Apart from the cluster being in equilibrium, sub- and superkeplerian situations could be in principle possible. Here we assume that the clusters are in virial equilibrium before gas expulsion. This probably justified, as for example the ONC is in a slightly supervirial state (Da Rio et al. 2014). The ONC can be regarded as a cluster in the early phases of gas expulsion, as the central part is nearly gas-free, whereas it is still embedded
in the non-central areas.

We used the Nbody6GPU code (Aarseth 2003, Nitadori \& Aarseth 2012) to model the dynamics of clusters after gas expulsion with a mass spectrum that represents a realistic IMF (Kroupa 2001). Many previous studies modelled such clusters with all stars having the same mass. However, Pfalzner \& Kaczmarek (2013a) showed that, at least for massive clusters ($M_c > 10^4$ \Msun), a full IMF has to be used, as otherwise the effect of gravitational interactions between cluster members are severely under-represented.   

In our simulations all stars were initially single, however, some binaries usually form during the simulations (Pfalzner \& Olczak 2007).   Starting with single stars is a standard procedure which significantly reduces the computation time. The stars were initially distributed according to a stellar density distribution equivalent to a W$_0$ = 9 King profile as observations show that the stellar density profiles of young clusters just before gas expulsion are best represented by King models with $W_0>$ 7 (for example, Hillenbrand \& Hartmann 1998).  We did not include mass segregation in the set up, because it is currently not clear whether the observed mass segregation in many young clusters is primordial or results from rapid dynamical evolution. 
More details of the model and the simulation method including a discussion of the approximations made can be found in
Pfalzner \& Kaczmarek (2013a,b).

We performed simulations for the model clusters with the parameters given by Table 1.
We assumed the same half-mass radius in all simulations  independent of the cluster mass, as investigation of the relation between the half-mass radius at the onset of gas expulsion point in that direction (Kroupa et al. 2010). The simulations were performed for each parameter set repeatedly with random seeds for the initial distribution (see Table 1), so that the errors in the cluster properties are generally $<$ 3-4\%.

In high-mass clusters ($M_c >$ a few 10$^3$ \Msun), due to the high number of massive stars all gas removal mechanism, like  photo-ionizing radiation, stellar winds and supernova explosions, are very effective, so that the gas is removed
in less than a crossing time (e.g. Melioli \& de Gouveia dal Pino 2006).  In this case, instead of explicitly simulating the gas expulsion process itself, the cluster can be treated as a stellar system out of virial equilibrium as described in  Bastian \& Goodwin (2006).

By contrast, low-mass clusters ($M_c <$ 1000 \Msun) do not sample the full IMF and therefore some of the feedback mechanisms are not or less effective. For example, supernovae explosions are not present and radiation pressure  becomes only important for clusters with masses in excess of 10$^{3.5}$\Msun\ (Cervino \& Luriana 2004, Krumholz \& Thompson 2012). As a consequence, gas expulsion probably happens on somewhat longer time scales, but cannot exceed 3 Myr, because no clusters older than 3 Myr are found that have a significant gas content. Thus for  lower mass clusters we treated not only the case of instantaneous gas expulsion (section 3), but also the case of a longer gas expulsion period (section 4).  In this case we model the gas expulsion as a background potential that decreases with time. 

The observations of cluster properties in the expansion phase after gas expulsion suffer from the problem that, unless proper motion data are available for all members, the bound and unbound cluster members occupy the same space for some time. Here we mimic this observational problem by taking into account all stars within 20 pc from the cluster centre until these two populations predominantly occupy different regions. The practical realisation of this method is described in more detail in
Pfalzner \& Kaczmarek (2013b). There we found that the mass-radius-age relation of the most massive clusters ($M_c > $ 10 $^4$ \Msun\ at ages $<$ 4 Myr) (Pfalzner 2009) in the solar neighbourhood is best reproduced with a mass of 1.5 - 4.5 $\times$ 10$^4$ \Msun, a half-mass radius in the range 1-3 pc and an SFE of 30\%. Here  we chose the simulation parameters for the most massive clusters (model A1) accordingly, meaning $N_c$= 30\,000, where $N:c$ is the number of stars, \mbox{$r_h$ =1.3 pc} and an SFE of 30\%.   Assuming an average stellar mass of 0.6 \Msun\ this corresponds to a cluster mass of $\sim$18\,000 \Msun.
The parameters of all our models, including the ones for lower mass clusters, are summarised in Table \ref{table:sim}.

\section{Results}

\begin{table}
\begin{center}
\begin{tabular}{lcrl}
ID & age & $M_{c}$ & $r_c$ \\[0.5ex]
\hline
GGD 12- 15$^a$ &4                & 73 & 0.615\\
NGC 2282$^b$   &5                 & 170 &0.8\\
NGC 2024$^c$    & 0.3           & 180 & 0.44 \\
IC 348$^d$         & 1.3 + 1.6  &160 & 0.5     \\
L 1641-C$^e$     & 0.1           & 27 & 0.155\\
NGC 2068$^f$   & 2               &110 & 0.43\\
NGC 2071$^f$    & 2               & 60 & 0.29\\
NGC 1333$^g$   & 1 &                79 & 0.245 \\
Mon R2$^h$        &0.8             &340 & 0.925\\
ONC                      & 1                & 1100 &1.9\\
CYg OB2$^i$        & 1-4    & 4.4   &  5.2 $^{}_{}$\\[0.5ex]
NGC 6611$^i$      & 1-5   & 4.4   &  5.9 $^{}_{}$\\[0.5ex]
NGC 2244$^i$      & 1-3    & 3.9   & 5.6  $^{}_{}$\\[0.5ex]
IC 1805$^i$         &  1-3    & 4.2   &  7.1 $^{}_{}$\\[0.5ex]
Ori Ib$^i$            & 1.7    & 3.6   &  6.3 $^{}_{}$ \\[0.5ex]
NGC 7380$^i$     & 2      & 3.8   &  6.5 $^{}_{}$ \\[0.5ex]
Ori Ic$^i$            & 4.6    & 3.8   & 12.5 $^{}_{}$  \\[0.5ex]
Ori Ia$^i$            & 11.4   & 3.7   & 16.6 $^{}_{}$  \\[0.5ex]
U Sco$^i$           & 5-11 & 3.5   & 14.2 $^{}_{}$\\[0.5ex]
Lower Cen-Crux$^i$  & 11-12 & 3.3 & 15.0 $^{}_{}$\\[0.5ex]
Upper Cen-Lup2$^i$ & 14-15 & 3.6 & 22.1 $^{}_{}$\\[0.5ex]
I Lac 2$^i$        & 12-16 & 3.4 & 20.7 \\[0.5ex]

\end{tabular}
\caption{ gives the properties of observed embedded clusters taken from Lada \&
Lada (2003). Here no errors are included as due to the embedded nature there
exists intrinsically a high uncertainity in the mass estimate. To these data the
cluster ages have been added form the following sources: $^a$Maaskant et al.
2011, $^b$Horner et al. 1997, $^c$Eisner \& Carpenter 2003, $^d$Herbig 1998, $^e$Allen \&
Davies 2008, $^f$Flaherty \& Muzerolle 2008, $^g$Raga et al. 2013, $^h$Andersen et al.
2008, $^i$Pfalzner 2009 and references therein.}
\label{table:obs}
\end{center}
\end{table}

\subsection{Average surface density}
\begin{figure}[t]
\includegraphics[width=0.45\textwidth]{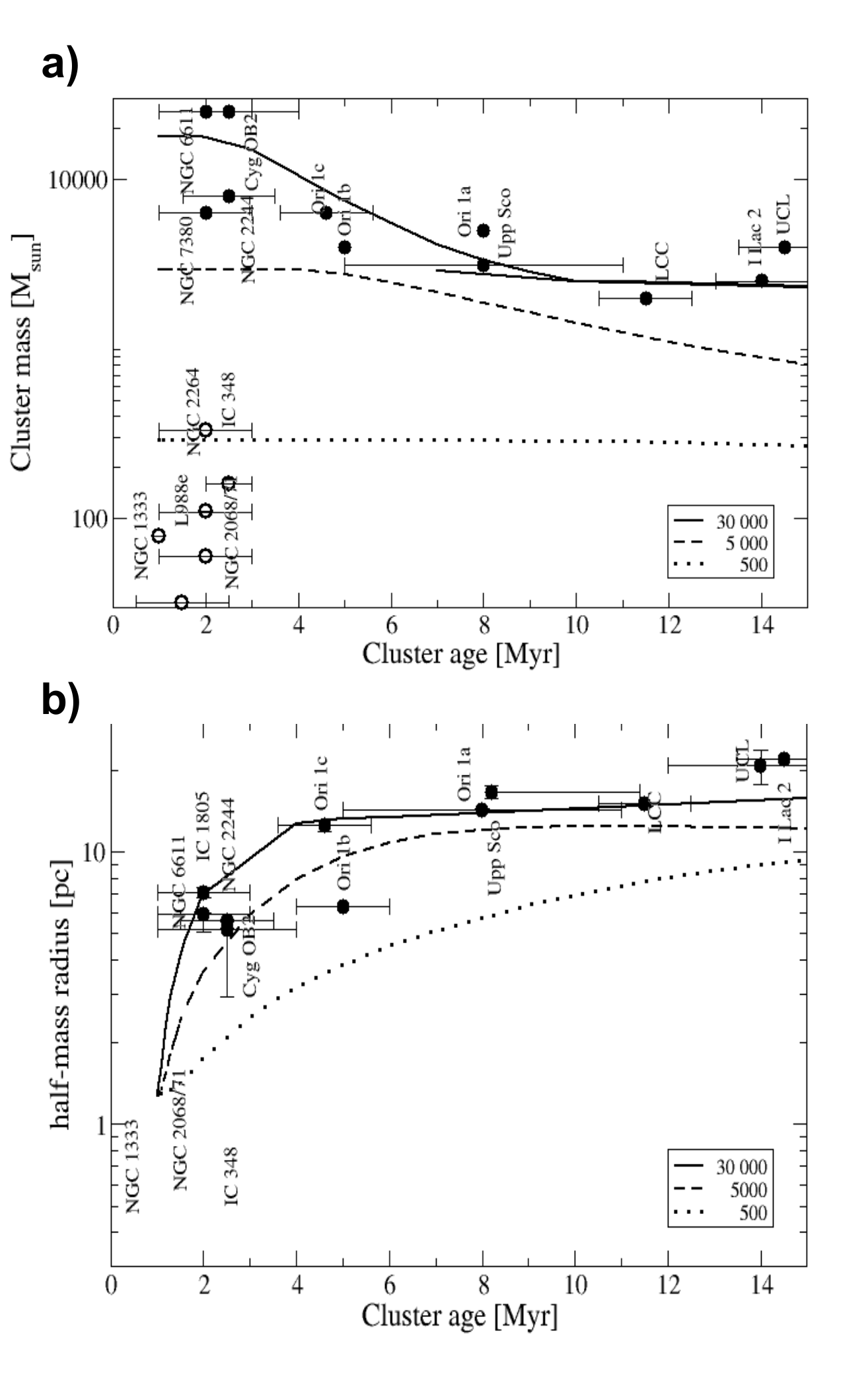}
\caption{Half-mass radii of the clusters as a function of their age. The open circles indicate clusters still containing large fractions of their natal gas (values from Lada \& Lada 2003), whereas the full circles represent clusters which are largely devoid of gas (values from Pfalzner 2009). The solid , dashed, and dotted lines represent our simulation results of the cluster development after gas expulsion for model cluster A1, A3, and A5, respectively. }
\label{fig:rad_age}
\end{figure}

 The circles in Figure \ref{fig:rad_age}  shows (a) the observed half-mass radii and (b) the mass development as a function of cluster age as given in Table 2. In addition, the error bars of the cluster age is indicated. Here observed means that no distinction between bound and unbound stars is made until they are spatially well separated. 
There are two distinct groups of clusters visible, young clusters (1-3Myr) that still contain considerable amounts of gas (open circles) and clusters that are largely gas free and cover a wider range of ages (2-14 Myr, full circles).
The embedded clusters are much smaller and of lower mass than the exposed clusters.   

The low-mass clusters indicated in Fig. 1 are all still embedded in their natal gas. If one only looks at the radius development (Fig. 1a) one could falsely assume that they are just younger versions of the exposed massive clusters. However, the observed embedded clusters have much lower masses than the exposed clusters (Fig. 1b). Although the embedded clusters still form stars and therefore gain stellar mass, their gas content does not suffice to increase the stellar mass by roughly two orders of magnitude. So why does one not observe older low-mass clusters? Most likely the young low-mass clusters either dissolve completely or fall below the detection limit at ages $>$ 3Myr (Lada \& Lada 2003, Pfalzner et al. 2014).

Here the aim is to model the temporal development of clusters of various masses to determine at what initial mass clusters fall later below the detection limit. In order to do that we need some knowledge about the cluster parameters at the end of the star formation process. We take the knowledge about the development of massive clusters as guideline. 

In Fig. 1 the solid line shows the result for model A1, which is representative for the development of a massive cluster. This is equivalent to the result of the model that in the parameter study of Pfalzner \& Kaczmarek (2013b) gave the best fit to the observational data. Here it was assumed that the clusters had an age of 1 Myr at the end of the gas expulsion process. Here the aim is to model the temporal development of lower mass clusters assuming that they have a similar SFE and initial radius. 

Figure 1 also shows the temporal development of the lower mass model clusters A3  (dashed line) and A5  (dotted line) consisting of 5000 and 500 stars, respectively. As expected from previous work (for example, Baumgardt \& Kroupa 2007), the lower mass clusters expand slower. Thus at the 15 Myr time span shown here their expansion is still ongoing. Eventually, the lowest mass cluster will expand to a half-mass radius of $\approx$ 12  pc, which is somewhat smaller than the final size of the massive clusters ($r_{hm} \approx $ 15 pc). The reason is that in massive clusters encounters induce additional member loss ($\approx$ 5\%), which the cluster compensates by further expansion (see Pfalzner \& Kaczmarek 2013a).

{\ bf The observed cluster properties are obtained without distinguishing between bound and unbound stars.} At the moment of gas expulsion a large fraction of the stars become unbound (40\%). However, over time even more stars become unbound because the unbound stars decrease the central potential even more. At the end of this process the remnant clusters contain only about 10\% of their initial membership. For the most massive clusters a clear distinction between the location of the bound and unbound stars becomes apparent at approximately 5 Myr. By contrast,  for our low-mass clusters (model A5) nearly all initial cluster members are still within a distance $<$ 20 pc from the cluster centre ( Fig.\ref{fig:rad_age}b).

Does this slower development mean we have been observing clusters with the same mass, radius etc. in completely different stages of their development? In other words, could we mistake a low-mass cluster consisting of bound and unbound stars for the remnants of a once massive cluster?

The mass development (Fig.\ref{fig:rad_age}b) shows that mistaking a cluster with initially 500 (model A5) stars for the remnant of one that contained 30 000 stars (model A1) is at no point in the development possible. Given the large uncertainty in mass and radius determination, deciding whether the observed clusters are remnants from model A1 or A2 is difficult for some of them if they are younger than 5 Myr. At older ages it is clear for most of them that they initially had a mass well above 10$^4$\Msun. Only for Upp Sco it remains unclear  whether it had initially a mass above 10$^4$\Msun\  as the age estimates range between 5 Myr and 11Myr.

The actual detectability of a cluster depends on the background surface density. Only clusters with a higher surface density than the background are detected unless proper motion measurements are available. Fig. \ref{fig:surface} shows the surface densities of the observed clusters and three of the model clusters 
(A1, A3, and A5). In panel a) the definition of surface density as the average number of stars divided by the area up to the half-mass radius of the cluster is used. The surface densities of all clusters decrease with cluster age, with those of the high-mass clusters being higher than those of the low-mass clusters. However, the differences are not very large.  Thus the differences in the average surface densities do not suggest that it would be difficult to detect  the remnants of the low-mass clusters.  

\subsection{Central surface density}

However, this picture is misleading.  In the solar neighbourhood clusters are usually first detected as a group of stars confined within a relatively small area (at most a few pc in diameter). The full extend in mass and radius of the clusters is usually only found later on by closer inspection. Therefore we have to compare the surface density of the central cluster area
rather than the one averaged over the entire cluster.

\begin{figure}[t]
\includegraphics[width=0.45\textwidth]{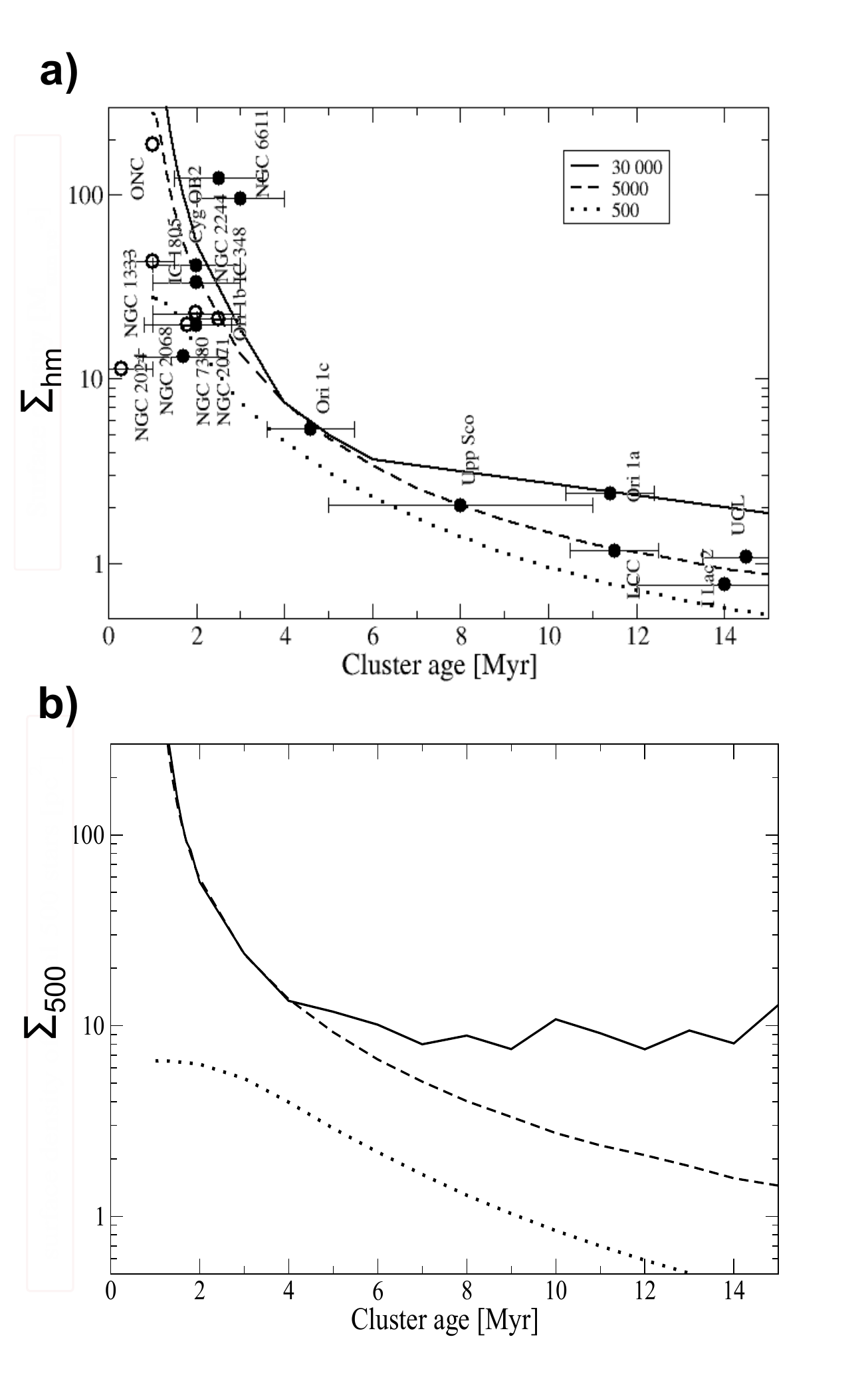}
\caption{Surface density for the same clusters as in Fig. 1, a) averaged at the half-mass radius and b) for the 500 most central stars.}
\label{fig:surface}
\end{figure}

Figure \ref{fig:surface} b) gives the average surface density of the 500 most central stars, $\Sigma_{500}$, as a function of cluster age for the same models as before. The differences in central surface density between the massive and the low mass clusters are much larger than before. At ages $>$ 4Myr  the most massive clusters have already more or less reached virial equilibrium. They show up to a hundred times larger  $\Sigma_{500}$ than those of low-mass clusters. This means that massive clusters in the age range would be easily detected against the background (at least in the solar neighbourhood), whereas low-mass clusters older than 4 Myr are easily missed. 

We chose the central 500 stars as measure, not because
this  is a typical number for first detection of clusters, but because the error bar for our model A1 is still tolerable. For model A3 and A5 statistical errors are less of a problem. The reason is two-fold: First, due to the higher computational demands the number of simulations is  considerably lower for the massive clusters with their large number of stars. Second, for the massive clusters (A1) the density is initially so high that scattering events are common. Model cluster A1 loses about 5\% of its initial mass by encounter processes (Pfalzner \& Kaczmarek 2013b). The actual encounter history in all the realisations of model A1 can vary considerably, with the effect of larger differences in the individual simulations.

Normally first detections of clusters can identify a few to about a hundred stars. Therefore a more realistic measure would be the surface density of the 50 most central stars $\Sigma_{50}$. This is what we will use in the following, when we have a closer look at the low-mass clusters.

\section{Dependence on expulsion time}

As mentioned before, the assumption of an instantaneous gas expulsion seems to be justified for
massive clusters, but is questionable for lower mass clusters. In order to investigate the dependence of the result on the duration of the gas expulsion phase, we performed simulations with the same initial parameters as model A5 but this time with a gas expulsion duration stretching over 2 Myr as well. 

\begin{figure}[t]
\includegraphics[width=0.49\textwidth]{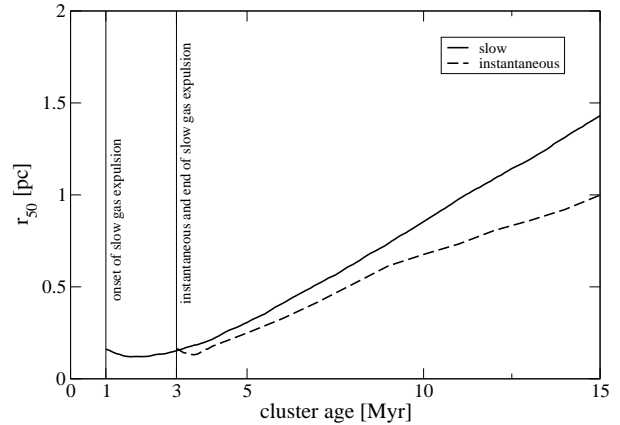}
\caption{Radius within which the 50 most central stars are contained as a function of cluster age for a cluster that contained 500 stars at the onset of gas expulsion. The solid line shows the case where gas expulsion sets in at 1 Myr and proceeds over the following 2 Myr, whereas the dashed line indicates the case of instantaneous gas expulsion at 3Myr.}
\label{fig:surface_low}
\end{figure}

Fig. \ref{fig:surface_low} shows the development of the cluster radius as a function of the cluster age for model A5 (dashed line) assuming  that the gas was instantaneously expelled at 1 Myr  and  for model B (solid line) where gas expulsion starts at 1 Myr and proceeds over a time period of 2 Myr. It can be seen that over the entire investigated time period the radius within which the 50 most central stars are contained $r_{50}$ of model cluster B is larger than that of model A5. 

Slower gas expulsion eventually leads to a higher bound mass of the cluster remnant. However, during the here investigated first 15 Myr the bound and unbound members are spatially indistinguishable, so that one would not observe any change in the system mass (see dotted line in Fig. 1). This means after gas expulsion the larger value of $r_{50}$ in model cluster B translates directly into  an even lower central surface density $\Sigma_{50}$ than model A5. This means low-mass clusters with slow gas expulsion are even less likely to be detected as such than those
with fast gas expulsion.

\section{Discussion}

\subsection{Star formation parameters}
In above simulations it was assumed that star formation proceeds in a similar way in low and high mass clusters. In principle, lower-mass clusters could form with a lower SFE. However, a lower SFE would lead to more mass loss and additional expansion resulting in an even lower surface density. Therefore our results assuming a SFE of 30\% can be regarded as upper limits of the surface density. In other words, if clusters below a certain mass and with a 30\% SFE are difficult to detect with current methods after gas expulsion, ones with lower SFE would be even harder to reveal. As a consequence the limit for the detection of clusters at ages $>$ 4Myr would move to an even higher value.

In order to test above model, predictions from these different models have to be tested against observations. Such predictions are provide here for the scenario of gas expulsion assuming a global star formation efficiency. In a follow-up paper we will do the same for a radially dependent SFE. These predictions can then be tested against the up-coming GAIA results to determine whether one of these models describes nature correctly.

\subsection{What will change with GAIA?}

 So far the old and intermediate-age open clusters NGC 6705, NGC 4815, and Trumpler 20 have been investigated (Magrini et al. 2014, Friel et al. 2014, Donati et al. 2014) in the GAIA-ESO survey. The only young cluster investigated so far is  $\gamma$ Velorum (Jeffries et al. 2014). These clusters were all previously known as surface density enhancements.

\begin{figure}[t]
\includegraphics[width=0.45\textwidth]{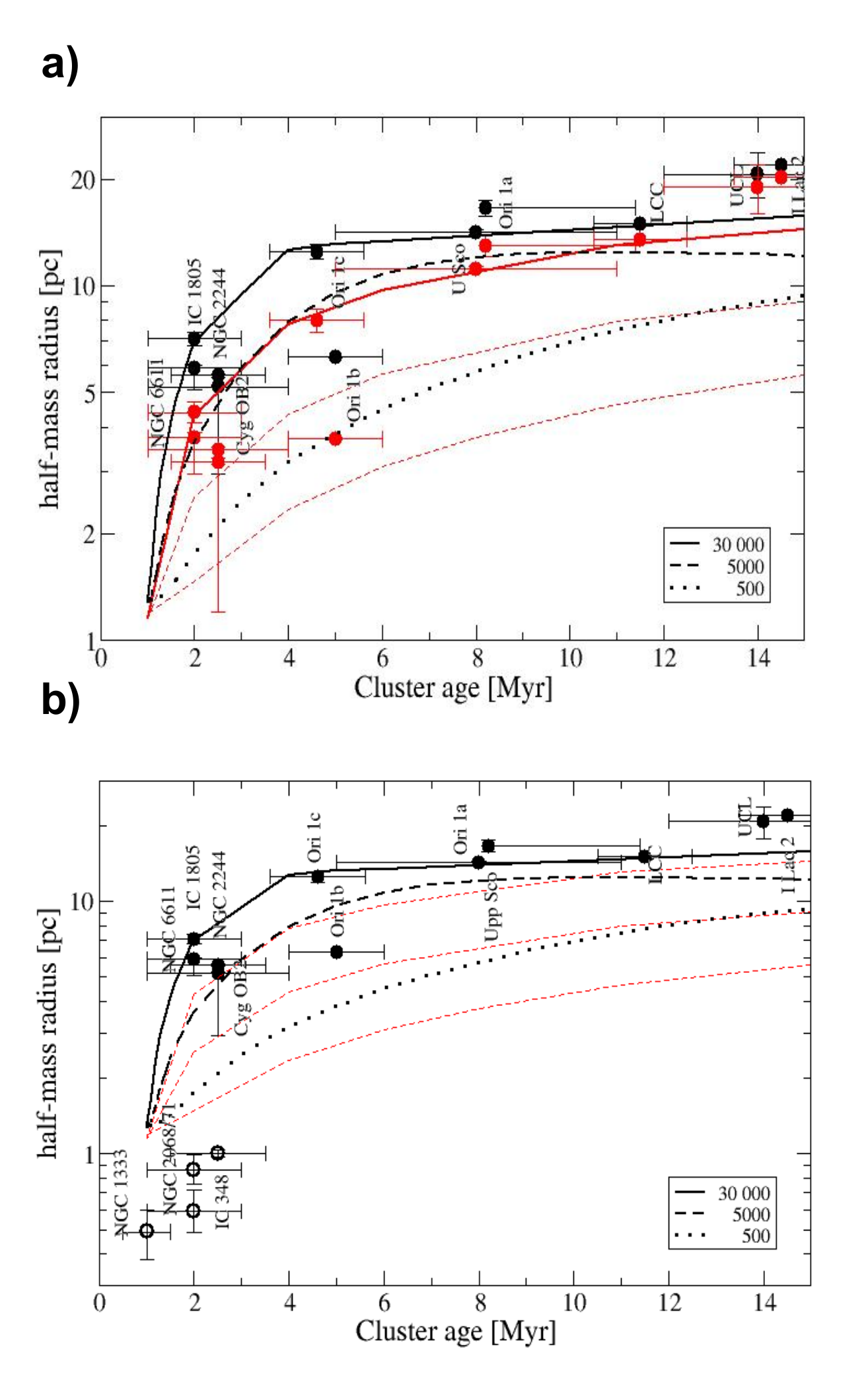}
\caption{Cluster mass as a function of cluster age for the same clusters as in Fig. 1. The black lines are the observed values, when the surface density enhancement against the background is used as measure. In this c ase no distinction between bound and unbound stars is possible. The red lines show the same values for the bound fraction alone. A colour version of this plot is available in the online version.}
\label{fig:milky}
\end{figure}

In surface density measurements there exists a problem in determining the cluster mass and radius just after gas expulsion. Apart from the actual bound stars, former cluster members are included in the mass and radius determination. These have become unbound but did not have enough time to leave the cluster area yet.  As a consequence the cluster membership as well as the cluster mass will usually be overestimated.

With GAIA a distinction between bound and unbound cluster members will be possible.   
Figure 4b shows the simulated bound cluster mass as a function of cluster age for our model clusters A1, A3, and A5 as red lines. Figure \ref{fig:rad_age}a) gives the corresponding cluster half-mass radii. The values that would be observed with surface density measurements are indicated as black line (same as Fig. 1).  In addition, the masses and half-mass radii  of specific clusters determined by surface density measurements are given (black circles). The red circles indicate the values we expect proper motion measurements should reveal for these clusters. These values are generally lower than those from surface density measurements as  only bound members are considered. 

The time when unbound stars are no longer mistaken for bound cluster members depends on the field of view. For the here assumed FOV of 40$\times$ 40 pc most unbound members of model cluster A1 have left the field of view at $\approx$ 5 Myr. Therefore
the difference is particularly pronounced for massive clusters in the age range 3-7 Myr. At later ages the difference between the two methods is much smaller. Thus one can expect that the  masses and half-mass radii of the older clusters UCL and LCC are representative for the remnant cluster. For the younger clusters the so far observed properties (black circles) can only be regarded as upper limits as they are contaminated with unbound members leaving the cluster. 

Figure 4 shows as well the expected mass and radius development for the lower mass cluster models A2 and A5. For both of them a considerable difference in observed and bound mass exists throughout the 15 Myr of development considered here. As these clusters expand more slowly and the bound members require more time to leave the cluster, in the early phases after  
gas expulsion both populations are identifiable as two distinct groups of stars with very different
velocity dispersions.

GAIA will as well be able to investigate the unbound population of the clusters. However, at later stages this extends to a very large distance from the cluster centre. For the most massive clusters this can extend up to $>$80 pc at an age of  20 Myr.

\section{Summary and conclusion}

This study determined which type of clusters is still detectable after gas expulsion at the end of star formation. Here it was assumed that the cluster had to have an enhanced surface density compared to the background in order to be detected as such.  Assuming a constant SFE throughout the cluster we find that 

\begin{itemize}
\item only clusters that contain more than 5000 $\pm$ 2000 stars before gas expulsion will usually be detectable at ages $>$ 4 Myr in areas typical for the solar neighbourhood. Clusters with lower membership will not be classified as such due to their low surface density. For a standard IMF this corresponds to clusters with a mass of approximately 3 $\times $ 10$^3$ \Msun. 

\item the number of cluster members and its spatial extend are overestimated in surface density based measurements as they can not distinguish between bound and unbound stars. During a transition phase the large fraction of stars that become unbound during the gas expulsion process (80-90\% of the initial cluster mass) are responsible for this overestimate. The duration of this phase depends on the initial cluster mass -
it last $\approx$ 5 Myr for clusters with initially 20 000 \Msun\ , whereas it extends beyond 20 Myr for clusters with a mass of 500 \Msun. 

\item in Fig. 4 we show  the predicted true masses and radii of specific clusters that will be detectable with proper motion measurements in the near future.

\end{itemize}

The temporal development of the properties of stars during the first 10 Myr of their development is nearly exclusively based on observations of these properties in clusters of this age range. These properties include the stellar properties like frequency of binaries and multiples, stellar rotation,  etc. and disc properties like the disc frequency, disc mass, disc size, accretion, dust size distribution etc. The advantage of using clusters to determine these properties is that a large number of fairly coeval stars are located close together. 

However, above described selection effect of only including  massive clusters at older ages  potentially has severe consequences.  One only obtains information for stars forming in massive clusters, whether this is applicable to lower mass clusters is an open question. (Pfalzner et al. 2014). This does not only concern the case of disc lifetimes (Pfalzner et al. 2014), but also the temporal development of other disc properties like accretion, dust growth etc. The same applies for the early development of binary properties which usually is determined in equally inhomogeneous samples. If low-mass clusters form with the same SFE as massive clusters, GAIA will be able to provide the older counterparts of the low-mass clusters. This will give then a homogeneous sample which will allow a
more consistent determination of the temporal development of all these properties.

\end{document}